# THERMIONIC SOURCES FOR ELECTRON COOLING AT IOTA[*]

M.K.Bossard[†], J. Brandt, Y.-K. Kim, S. Kladov, The University of Chicago, Chicago, USA
N. Banerjee, B. Cathey, G. Stancari, Fermi National Accelerator Laboratory, Batavia, USA
S. Nagaitsev, Thomas Jefferson National Accelerator Facility, Newport News, USA

*Abstract*

We are designing and fabricating two new thermionic sources of magnetized electrons for use in the electron lens project in the Integrable Optics Test Accelerator (IOTA) at Fermilab. These electron sources will be used for cooling 2.5 MeV protons. One source will be used to compensate for emittance growth due to Intra Beam Scattering in experiments with weak space-charge, while the other source will be used to research the interplay between electron cooling and intense space charge. In this paper we present the progress made so far and the upcoming steps for the thermionic electron sources.

## INTRODUCTION

Throughout history, electron sources have played an important role in various fields, examples including the development of vacuum-tube technology, modern electronic circuits, microscopy, and particle accelerators [1]. Towards these and other applications, electron sources produce electron emission, which refers to the process where electrons are emitted from a material when their energies exceed the material's work-function [2]. Various physical stimuli, such as thermal-energy, photons, ion or electron bombardment, and high magnetic fields can supply the energy required for electron emission [3]. Our project specifically focuses on thermionic electron emission, which uses thermal-energy as the physical stimuli. Thermionic electron sources are commonly used within electron lenses, which are flexible instruments for both particle accelerator operations and beam physics research [4].

Electron lenses employ magnetically confined, low-energy electron beams, which overlap with recirculating beams in a storage ring. [4] The electron beams that are produced by electron lenses are highly stable within a solenoidal magnetic field. The electron beams interact with the recirculating charged particles by means of electromagnetic fields, allowing for changes in the distribution of the recirculating particles in phase-space. Electron lenses have many uses, including space-charge compensation, halo reduction [7], beam-beam compensation [5, 6], and electron cooling.

The Integrable Optics Test Accelerator (IOTA) at Fermilab is a re-configurable 40 m storage ring [8, 9], serving as a research facility for intense beams, including the areas such as space-charge [11], Non-linear Integrable Optics (NIO) [10], beam cooling [4], and single electron storage [13]. It is capable of circulating both protons and elec-trons at kinetic energies up to 2.5 MeV and 150 MeV, respectively [11]. Figure 1 shows the layout of IOTA, including the location of the electron cooler within the ring. Protons are injected in section A and circulate clockwise, co-propagating with the electrons in section DR.

___________
[*] WORK SUPPORTED BY THE UNIVERSITY OF CHICAGO AND FERMI NATIONAL ACCELERATOR LABORATORY
[†] mbossard@uchicago.edu

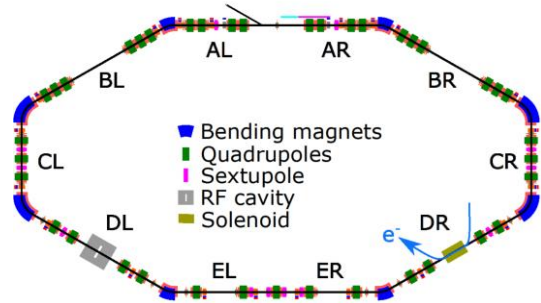

Figure 1: Schematic of the Integrable Optics Test Accelerator (IOTA). The ring is separated into sections A-E. The blue arrow in section DR represents the electron beam path.

Our project aims to design, test, and commission two thermionic electron sources at IOTA. The first source, called the weak source, will be used for cooling proton beams with relatively small currents and performing beam manipulations relevant to other experiments in IOTA. The second source, called the strong source, will play a role in investigating the control and dynamics of proton beams with electron cooling under intense space-charge [11].

In the following sections, we will discuss electron cooling in IOTA, the designs of the cooling sources, and the development of the test stand for such sources. We then finally present our next steps and conclude [12].

## ELECTRON COOLING

When viewed from the center of mass frame of a recirculating ion beam, the ions have random transverse velocities, which results in a large emittance and high thermal energy. This thermal energy is increased by non-linear forces and interactions within the beam-line, such as intra-beam scattering, space-charge effects, and external non-linear magnetic fields along the accelerator path. A cooling method must be used in order to decrease this thermal energy of the ion beam.

As an ion beam recirculates through the accelerator, its thermal energy increases. To reduce this, electron cooling can be used. Electron cooling involves exchanging thermal energy between a beam of ions and a co-propagating beam of electrons. In this methode, the ion beam is periodically mixed with a bath of cold, low-energy electrons. The elec-

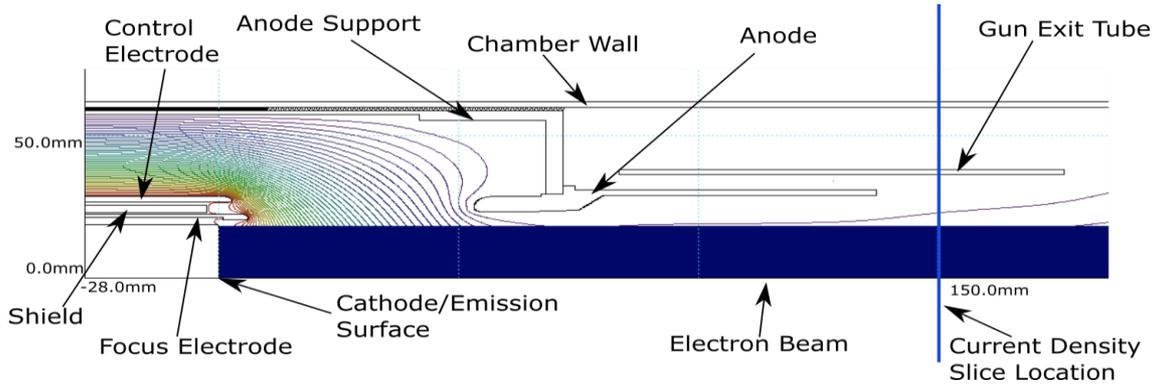

Figure 2: Present design of the strong thermionic source for electron.

Table 1: Electron Cooler Parameters for IOTA

| Parameter | Values | | Unit |
|---|---|---|---|
| **Proton parameters** | | | |
| RMS Size ($\sigma_{b,x,y}$) | 3.22, 2.71 | | mm |
| **Main solenoid parameters** | | | |
| Magnetic field ($B_\parallel$) | 0.05 - 0.5 | | T |
| Length ($l_{cooler}$) | 0.7 | | m |
| Flatness ($B_\perp/B_\parallel$) | $2 \times 10^{-4}$ | | |
| **Electron parameters** | | | |
| Kinetic energy ($K_e$) | 1.36 | | keV |
| Temporal Profile | DC | | |
| Transverse Profile | Flat | | |
| Source temp. ($T_{cath}$) | 1400 | | K |
| **Source parameters** | | | |
| Parameter | Weak | Strong | |
| Current ($I_e$) | 1.7 | 80.3 | mA |
| Radius ($a$) | 14.1 | 18 | mm |
| Magnetic Field ($B_\parallel$) | 0.4 | 0.1 | T |
| $\tau_{cool,x,y,s}$ | 7.6, 6.5, 5.3 | 2.5, 2.4, 5.3 | s |

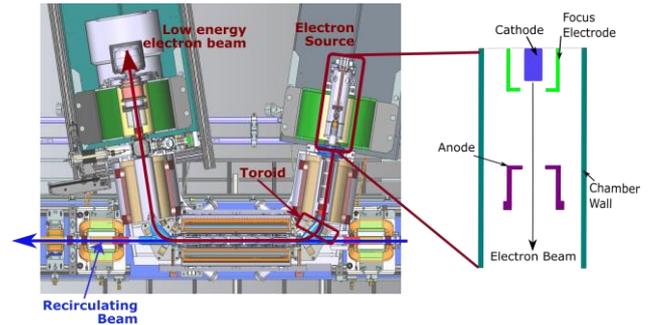

Figure 3: Electron lens setup at IOTA.

## ELECTRON SOURCE DESIGN

We have been utilizing the simulation software TRAK [16] to assist in the engineering of designs [12] for the thermionic electron sources. The design for the strong thermionic source in TRAK can be seen in Fig. 2. The main features of the electron source are depicted, where the entire system is cylindrically symmetric. The dark blue lines depict electron trajectories and the color gradient shows the electric field equipotentials, including beam effects. We use a cylindrical barium impregnated tungsten dispenser cathode with a flat emission surface. The focus electrode encircles the cathode and ensures that the emitting electrons do not diverge as they are emitted from the cathode. The anode is located further away and supplies an accelerating potential, forcing the electrons down the beam line. The entire system is simulated in a vacuum chamber with a magnetic field of 0.1 T for the strong source. Alternatively, the magnetic field for the weak source is 0.4T, to allow for beam expansion in the main ring. These magnetic fields constrain the motion of the electrons into helices, forcing the beam to remain collimated rather than diverging throughout its entire path.

The most relevant beam parameters from the source output include the total current emitted, the radius of the beam, the maximum electric field, and the transverse beam current density distribution. We have explored optimizing the focus electrode and anode positions to obtain a flat beam distri-trons move at the same average longitudinal velocity as the ions, and the beams undergo Coulomb scattering, resulting in a reduction of the thermal energy of the ion beam. [14,15].

Figure 3 displays a close-up of the DO section of the IOTA ring, with the location of the electron cooler depicted, along with the path taken by the electron beam. Solenoids are utilized throughout the electron beam's propagation path to maintain its magnetization. The schematic of an electron source is also illustrated, with the electrodes providing the required potential differences and heat for electron emission [12].

We will create two electron sources for cooling, a weak electron source and a strong electron source, with their parameters listed in the lower portion of Table 1. Their design process will be further discussed the next section.

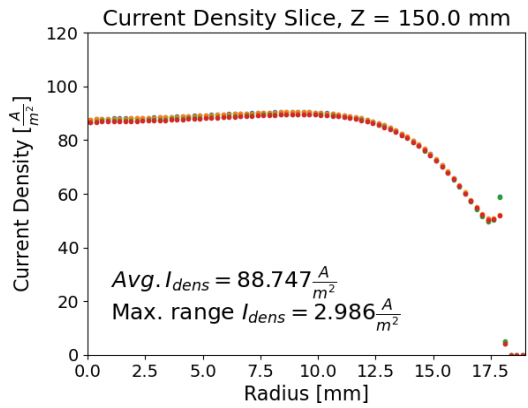

Figure 4: Beam distribution at 150 mm from the cathode for the strong thermionic source

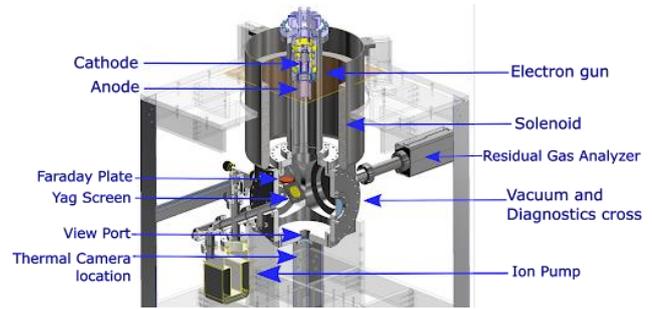

Figure 5: Test stand setup with an electron source in place.

bution and appropriate current. This was done through the python-based optimizer, Xopt [18]. Specifically, we used the Non-dominated Sorting Genetic Algorithm within Xopt. This optimization method resulted in the final strong source parameters seen in Table 1. The resulting current density distribution for the strong source at 150 mm from the cathode can be seen in Fig. 4. The beam flatness condition is only required to a radius up to 12mm for the strong source, and as seen in this figure, the current density up to 12mm is a flat distribution with deviations within 10% of the mean current density. Furthermore, all simulation results have reported a maximum electric field ≤ 120 kV/m, an order of magnitude below the limit of 1000 kV/m, where arcing would be a relevant issue. Therefore there is very little concern about arcing in this geometry.

Furthermore, the simple source design is now in progress. The most relevant simulation results for the simple source are reported in Table 1. This source has no anode, as the lower limit of emission current is determined by the grounded chamber wall. Even without an anode, the simulation results show difficulty with reducing the current density down to the desired value for the simple source. The goal for the current density is ≤ 0.84 $A/m^2$, where the simulation results are presenting 1.92 $A/m^2$. Other options, such as reducing the duty cycle of the electron source, will be explored to reach the desired cooling rates for the simple electron source.

Before commissioning the new electron sources at IOTA, we need to verify their operation at the University of Chicago. The design of a test stand for such verification is shown in Fig. 5. For more information on the test stand, please refer to S.Kladov's contribution for this conference [19].

## CONCLUSIONS AND NEXT STEPS

We are creating electron sources to cool proton beams and research the influence on electron cooling ion space-charge forces at high-intensity limits. To generate appropriate electron beams, we are designing and simulating two electron sources. One is a simple source be used for other experiments at IOTA, where the other is a strong source to be used to study the interplay between electron cooling and space-charge in the high-intensity limit. The design for the strong source has been completed using the TRAK simulation package and Xopt's CNSGA optimization method. We are currently working towards the simple source design. A test stand at UChicago has been built to test these and other sources.

In the future, to determine the simple electron source design, the geometries of the source will be altered through optimization. Once both of these sources are designed and their operations confirmed in TRAK, they will be manufactured, tested on the test stand, and finally commissioned in the IOTA ring.

## ACKNOWLDGEMENTS

We would like to thank the teams at the University of Chicago, Fermilab, and the IOTA/FAST collaboration for their support. This research is funded by the NSF Graduate Research Fellowships Program (GRFP). The project is a collaboration between the University of Chicago and Fermilab. This manuscript has been authored by Fermi Research Alliance, LLC under Contract No. DE-AC02-07CH11359 with the U.S. Department of Energy, Office of Science, Office of High Energy Physics.